\documentclass{article}
\usepackage{ijcai19}

\usepackage{times,soul,url,graphicx,amsmath,amsfonts,booktabs}
\usepackage[hidelinks]{hyperref}
\usepackage[utf8]{inputenc}
\usepackage[small]{caption}
\usepackage{algorithm}
\usepackage{algorithmic}
\usepackage{amsthm,mathrsfs}
\newtheorem{assumption}{\textit{Assumption}}
\newtheorem{definition}{\textit{Definition}}
\newtheorem{lemma}{\textit{Lemma}}
\newtheorem{theorem}{\textit{Theorem}}

\title{Estimating Individual Advertising Effect in E-Commerce}

\author{
Hao Liu$^{1,3}$\footnote{Contact Author}\and
Yunze Li$^2$\and
Qinyu Cao$^2$\and
Guang Qiu$^2$\And
Jiming Chen$^{1,3}$\\
\affiliations
$^1$College of Control Science and Engineering, Zhejiang University\\
$^2$Alibaba Group\\
$^3$Alibaba-Zhejiang University Joint Institute of Frontier Technologies\\
\emails
\{vectorliu, cjm\}@zju.edu.cn,
\{yunze.lyz, qingyu.cqy, guang.qiug\}@alibaba-inc.com
}

\begin{document}

\maketitle

\begin{abstract}
Online advertising has been the major monetization approach for Internet companies. Advertisers invest budgets to bid for real-time impressions to gain direct and indirect returns. Existing works have been concentrating on optimizing direct returns brought by advertising traffic. However, indirect returns induced by advertising traffic such as influencing the online organic traffic and offline mouth-to-mouth marketing provide extra significant motivation to advertisers. Modeling and quantization of causal effects between the overall advertising return and budget enable the advertisers to spend their money more judiciously. In this paper, we model the overall return as individual advertising effect in causal inference with multiple treatments and bound the expected estimation error with learnable factual loss and distance of treatment-specific context distributions. Accordingly, a representation and hypothesis network is used to minimize the loss bound. We apply the learned causal effect in the online bidding engine of an industry-level sponsored search system. Online experiments show that the causal inference based bidding outperforms the existing online bidding algorithm.
\end{abstract}

\section{Introduction}\label{sec:introduction}

The last two decades have seen the prosperity of e-commerce. Taking Taobao as an example, as the biggest e-commerce marketplace in China \cite{edquid17}, Taobao search service covers over $300$ million consumers each day, bringing daily $10$ billion search queries and subsequent page views (PVs), providing advertisers sufficient opportunities to promote their commodities online \footnote{https://zhitongche.taobao.com}.

In sponsored search advertising, advertisers bid for keywords associated with their commodities (ADs \footnote{A commodity is also called an AD in Taobao, which represents both the commodity and the associated advertisement.}) and pay the platform when consumers land their commodity/store homepage by clicking the advertisement (Pay-Per-Click, PPC). The payment equals the minimum bid price required to keep the advertising slot in the real-time competition \cite{wilkens2017gsp}. The returns of advertising can be summarized in two aspects. The \emph{direct returns} of advertising are the impressions, clicks and conversions occurred upon the advertising PVs. Meanwhile, advertising also yields even more valuable \emph{indirect returns} by connecting with wider online audience, thus impressing more audience via unobserved social interactions. Additionally, in e-commerce platform like Taobao, there is a ranking index called ``sales volume'' for organic search traffics, which reflects the purchasing popularity of the commodity among its peers and has been an important shopping guideline for consumers. In this way advertisers can accumulate sales volume via advertising PVs to gain more exposures in gigantic organic search PVs.

The direct and indirect returns motivate advertisers to invest advertising budget to prosper their online business. Despite its significance, however, to the best of our knowledge, existing work mostly focus on optimizing direct returns \cite{zhang2016optimal,zhu2017optimized}. This might be caused by the fact that direct returns are seamlessly observable in the closed-loop e-commercial platform like Taobao. Meanwhile, there are so many factors leading to the \emph{overall returns}, making it intractable to quantify the indirect returns attributed to advertising. Nonetheless, the ability of inferring the overall advertising effect including both direct and indirect returns provide advertisers the opportunities to allocate their advertising budget more wisely in the product life cycle.

E-commercial advertisers are eager to know the growth of advertising returns if they invest more budget via a specified advertising channel. Specifically, in the PPC advertising, the cost is equivalent to the number of clicks occurred in advertising PVs given a relatively stable payment per click. And the overall advertising returns can be observed as the number of total clicks of the advertising AD accumulated in all the online channels. Therefore, we want to infer the \emph{individual advertising effect (IAE)} via predicting the incremental number of all-channel clicks in a period under the intervention of advertising clicks \footnote{Numerical analysis also shows that the Pearson's coefficient between the number of advertising clicks and all-channel clicks is approximately $25\%$ larger than that between advertising impressions and all-channel clicks. We hide the detailed coefficient value due to commercial secrets.}. 

The problem of inferring IAE resembles the estimation of \emph{individual treatment effect (ITE)} in the field of \emph{causal inference} or learning from \emph{observational data} \cite{rubin2005causal}. In causal inference, we only have observational data which contains the past actions, their outcomes and possibly more context. However, we do not know the mechanism which gave rise to the action. In the scenario of advertising, the context might correspond to the features representing the current status of the AD, while the action and outcome are the number of advertising clicks and all-channel clicks (containing all direct and direct returns), respectively. The key difference between IAE and ITE is that actions of the latter are binary or categorical, but those of the former might be continuous and transitive. Furthermore, for any specified context, there exists only one exact action (acquire specific number of advertising clicks) in the data. We can never know exactly the potential advertising outcome if it applies a different action in exactly the same context. Besides, the observed advertising outcomes can be influenced by a lot of factors including online sources such as in or out of Taobao recommendation/organic search, and offline mouth-to-mouth marketing by the audience etc, which is similar to the \emph{confounding} factor in classic causal inference. Since the effect of advertising is accumulating in the whole-time horizon, we assume that the context together with the action contain all the necessary information to determine the outcome, i.e., the ``no-hidden confounding'' assumption holds in the analysis.

In this paper we model the causal effect between advertising cost and returns via a formal definition of IAE. We propose a representation network and a hypothesis network combined to predict the individual advertising effect referring \cite{shalit2017estimating}. Different from binary or categorical treatments, advertising treatments are continuous and transitive. Relying on this property, we derive a rigorous theoretical upper bound of the expected IAE estimation error by way of a learnable factual regression loss and the distance of context distribution among different treatments. Then the network is trained to minimize the derived theoretical upper bound. Furthermore, we derive a time-varying factor called \emph{leverage rate} ($lvr$) based on IAE to reflect the AD-level potential to lever the overall advertising returns. The learned $lvr$ is used in the online bidding engine to achieve better overall advertising performance in Taobao sponsored search. The contributions of this paper can be summarized as follows: 
\begin{enumerate}
  \item We model the problem of predicting the overall advertising return in the framework of causal inference. In this framework, the formal definition of individual advertising effect is given.
  \item We derive a general theoretical upper bound on the expected IAE estimation error in advertising scenarios with multiple continuous and transitive treatments. Subsequently, a representation and hypothesis network is learned to predict IAE. 
  \item IAE-induced $lvr$ is integrated in the online bidding engine, which yields better overall advertising returns compared with the existing bidding engine in Taobao sponsored search.
\end{enumerate}


\section{Related Work}\label{sec:related_work}

When considering only direct advertising returns, estimating individual advertising effect has been investigated in both ex ante and ex post way. The key of the ex ante estimation lies in three separate models of predicting the winning rate of specific bids, click-through-rate and conversion-rate of the advertising PV \cite{zhang2016optimal,zhu2017optimized}. Meanwhile, attribution modeling \cite{dalessandro2012causally,diemert2017attribution} corresponds to the ex post estimation of advertising effect, i.e. attributing the later conversion to the previous customer-commodity/store contact. All these works ignore the indirect returns brought by the advertising PVs.

Causal inference has already been used in complex real-world ad-placement systems \cite{bottou2013counter}. In the scenario of estimating individual advertising effect, given a context (AD), a naive way might be direct least square regression to fit the advertising effect, either taking the number of advertising clicks as a feature or separately fit each action. However, such estimation might be biased due to the fact that different contexts should have priority of choosing specific actions in the dataset. To alleviate the bias, propensity score, which characterizes the probability vector of choosing specific actions, \cite{austin2011introduction} is used to evaluate the similarity of two contexts. Therefore, counterfactual samples can be constructed by comparing the propensity scores via various approaches such as nearest neighbor matching \cite{lopez2017estimation}. Besides propensity score, various methods such as random forests\cite{wager2017estimation,athey2016recursive} and expensive random control trials \cite{taddy2016nonparametric,peysakhovich2016combining} are also used to tackle the binary treatment causal inference.   

Recently deep representation is also used to encode the contexts. Atan \emph{et al.} proposed an auto encoder-decoder network to represent the raw context, to ensure that the propensity score vector of the mapped contexts is similar, therefore removing the selection bias \cite{atan2018deep}. Johansson \emph{et al.} also designed a deep representation network to embed the original contexts, to guarantee that the distribution of contexts after the representation is similar between two different treatments, as well as the small regression loss \cite{johansson2016learning}. In its later version, a theoretical error bound on the expected ITE is given to yield a more rigorous estimation algorithm \cite{shalit2017estimating}. Deep models prove to be advanced but when faced with the multiple treatments, theoretical error bound is non-trivial. In this paper we design a similar network structure as \cite{shalit2017estimating} but derive our theoretical upper bound considering the continuous action space in the advertising scenario.

Back to the bidding application, perhaps the most relevant work is lift-based bidding proposed by Xu \emph{et al}. By predicting the ex ante and ex post click-through-rate of an advertising impression, the bid price is adjusted to be proportional to the lift \cite{xu2016lift}. However, we point out that the observed outcome might also ignore the abundant indirect returns.

\section{Individual Advertising Effect Formalization}\label{sec:lvr_definition}

We adopt the Rubin-Neyman potential outcomes modeling framework \cite{rubin2005causal} in causal inference but tailor it for the e-commerce advertising scenario. Let $\mathcal{X} \subset \mathbb{R}^d$ be the set of contexts, $\mathcal{T} = \{T_1,...,T_n\}$ the $n$-action (also known as treatment or intervention) set, $\mathcal{Y} \subset \mathbb{R}$ the observed \emph{overall} performance index. For each context $x \in \mathcal{X}$, there is a treatment assignment $T \in \mathcal{T}$ and with $n$ potential outcomes, $Y_{T_1}, Y_{T_2}, ..., Y_{T_n} \in \mathcal{Y}$. The samples we have can be denoted as $\{x_i, t_i, y_i\}_{i=1}^{N}$, where $y_i = Y_{t_i}$. We do not observe any of the other potential outcomes (i.e., $Y_{T}$ for $T \neq t_i$). 

In e-commercial advertising, context $x$ can be features representing the status of an AD in the beginning of the day. Treatment $T$ refers to the number of clicks acquired from the advertising PVs during the day, while potential outcome $y$ might be observed as the overall whole-site clicks obtained by the same AD until the end of the day. Apparently, the potential outcome can be influenced by a lot of factors including online channels such as in or out of Taobao recommendation/organic search, and offline mouth-to-mouth marketing by extroverted audience etc. Specifically, let $T_i = i-1, i=1,...,n$, where $n-1$ can be interpreted as the context-specific largest possible advertising clicks in a day. Note that we restrict the advertising effects $y$ to be happened in the same day, but ignores the persisting effects in the far future. This naturally coincides with the advertising logic that advertisers are accustomed to adjusting the budget of an AD day by day. We can also alleviate the influence of persisting dependency by following the ``strong ignorability'' assumption in causal inference.

\begin{assumption}
(\emph{Strong Ignorability}) $Y_{T_1}, Y_{T_2}, ..., Y_{T_n} \perp T|x$, which means that, given a context $x$, the potential outcome is independent of the treatment assignment. 
\end{assumption}
Strong ignorability assumption also ensures that there is a positive probability of choosing any action in each context $x$.

\begin{definition}
(Individual Advertising Effect, IAE) The IAE of context $x$ from treatment $T_i$ to $T_j$ can be defined as:
\begin{equation}\label{eqn:iae_definiton}
\alpha_{i,j}(x) = m_j(x) - m_i(x), \forall i,j=1,...,n,
\end{equation}
where $m_i(x) = \mathbb{E}[Y_{T_i}|x],\forall i,j=1,...,n$.
\end{definition}

For $n$ treatments, we can obtain an antisymmetric matrix $\alpha(x) = [\alpha_{i,j}] \subset \mathbb{R}^{n \times n}$ which corresponds to the IAE in context $x$. Matrix $\alpha(x)$ has the following properties:
\begin{enumerate}
\item [$\bullet$] \emph{Antisymmetric:} $\alpha_{i,j} = - \alpha_{j,i}$;
\item [$\bullet$] \emph{Monotonicity:} $\alpha_{i,j} \le \alpha_{i,k}, \forall j \le k, i=1,...,n$; $\alpha_{i,j} \ge \alpha_{k,j}, \forall i \le k, j=1,...,n$;
\item [$\bullet$] \emph{Zero-Diagonal:} $\alpha_{i,i} = 0, \forall i=1,..,n$;
\item [$\bullet$] \emph{Transitivity:} $\alpha_{i,j} + \alpha_{j,k} = \alpha_{i,k}, \forall i,j,k=1,...,n$.
\end{enumerate}

In causal inference, IAE is analogous to individual treatment effect. However, different from classical causal inference with only two interventions of either treatment or non-treatment, multiple actions are available in IAE. Since the action and outcome are accumulated in a day, there is some kind of ambiguity of advertising effect with clicks assigned to different time slots. Therefore, we take the expectation in the right-hand side of Eqn. (\ref{eqn:iae_definiton}) to eliminate the ambiguity. In this sense, IAE is the average advertising effect and should be useful among different ADs. 

To learn IAE, we further define a representation function $\Phi: \mathcal{X} \to \mathcal{R}$ where $\mathcal{R}$ is the representation space. Let $h: \mathcal{R} \times \mathcal{T} \to \mathcal{Y}$ be a hypothesis function which yields the outcome. Putting it together, we denote $f(x, T) = h(\Phi(x), T)$.

\begin{definition}
Given a hypothesis $f$, the IAE estimation for context $x$ is:
\begin{equation}
\hat{\alpha}_{i,j}^{f}(x) = f(x, T_j) - f(x, T_i).
\end{equation}
\end{definition}
With a little abuse of notation, we will omit the superscript $f$ and write it as $\hat{\alpha}_{i,j}(x)$ without confusion.

\begin{definition}
The IAE estimation error of treatment pairs $(T_i,T_j)$ satisfies that:
\begin{equation}
\tau_{i,j}(x) = \hat{\alpha}_{i,j}(x) - \alpha_{i,j}(x). 
\end{equation}
\end{definition}

\begin{definition}
The expected \emph{Precision in Estimation of Heterogeneous Effect (PEHE)} \cite{shalit2017estimating} loss of $f$ is:
\begin{equation}\label{eqn:pehe_loss}
\epsilon_{\text{PEHE}}(f) = \frac{1}{n(n-1)}\sum\limits_{i=1}^{n}{\sum\limits_{j=1,j\neq i}^{n}{\int_{\mathcal{X}}{\tau_{i,j}^2(x)p(x)dx}}}.
\end{equation}
\end{definition}

For completeness, we also give the notion of \emph{Integral Probability Metric} (IPM), which is a class of distance metrics between probability distributions \cite{shalit2017estimating}. For two probability density functions $p, q$ defined over $\mathcal{S} \subset \mathbb{R}^d$, and for a function family $G$ as $g: \mathcal{S} \to \mathbb{R}$, it holds that
\begin{equation*}
\text{IPM}_G(p,q) := \sup\limits_{g \in G}|\int_{\mathcal{S}}{g(s)(p(s) - q(s))ds}|.
\end{equation*} 


\section{Learning to Infer Individual Advertising Effect}\label{sec:learn_lvr}
\begin{figure}
\includegraphics[width=3.2in]{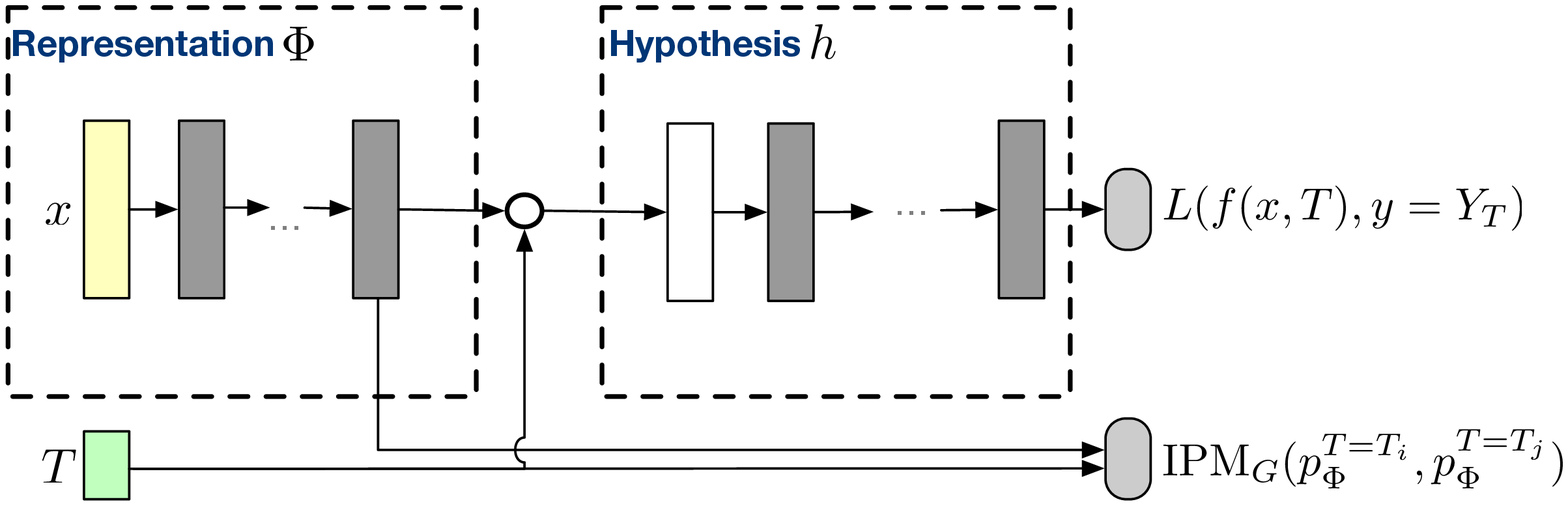}
\caption{Neural network architecture for IAE estimation. $L$ is a loss function and $\Phi$ is a representation of the original context $x$. $h$ represents the hypothesis and $f$ denotes the complete function.}\label{fig:lvr_model}
\end{figure}

The key of learning IAE lies in minimizing the PEHE loss in Eqn. (\ref{eqn:pehe_loss}). The idea is similar as \cite{shalit2017estimating}. Firstly, we map the original context into the representation space $\mathcal{R}$ via $\Phi$. In the new space, we denote the probability density function of representation space $\mathcal{R}$ given treatment $T$ as $p_{\Phi}^{T}$. To remove the selection bias in the original space, the idea is that the distance between $\{p_{\Phi}^{T_i}\}_{i=1}^{n}$ should be as small as possible, which is guaranteed by the representation network. Given the similar context distribution in the representation space, the hypothesis network should try to minimize the regression loss of fitting the advertising return. The neural network architecture is displayed in Fig. \ref{fig:lvr_model}.

The architecture resembles that in \cite{shalit2017estimating}. However, the key differences lies in two aspects. Firstly, the hypothesis network in that of \cite{shalit2017estimating} are separate for treatment/non-treatment. In the advertising scenario, the treatments can be seen as continuous actions and should be generalizable, therefore different treatments share the same hypothesis network. Secondly, IPM for binary treatments are straightforward while it is not obvious for multiple treatments. We simplify the IPM term based on the transitive property of treatment effects. In the following part we will first give the theoretical upper error bound of PEHE error and elaborate the detailed IPM we use, followed by the description of the IAE estimation algorithm. 

\subsection{Loss Error bound}\label{sec:loss_bound}

Before analyzing our main result, we first give a lemma considering the binary treatment case.
\begin{lemma}\label{lemma:binary_loss}
\cite{shalit2017estimating} Let $\Phi: \mathcal{X} \to \mathcal{R}$ be a one-to-one representation function with inverse $\Psi$. Let $h: \mathcal{R} \times \{T_i, T_j\} \to \mathcal{Y}$ be an hypothesis. Assume there exists a constant $B_{\Phi}$ such that for $T \in \{T_i, T_j\}$, the per-unit expected loss functions $\ell_{h,\Phi}(\Psi(r),T)$ obey $\frac{1}{B_{\Phi}}\cdot \ell_{h,\Phi}(\Psi(r),T) \in G$, where $\ell_{h,\Phi}(x,T) = \int_{\mathcal{Y}}{L(h(\Phi(x), T), Y_T)p(Y_T|x)dY_T}$. Assuming that the loss $L$ is the squared loss, we have that
\begin{equation*}
\begin{aligned}
& \int_{\mathcal{X}}{\tau^2_{i,j}(x)p(x)dx} \le \\
& 2(\epsilon_{F}^{T=T_i}(h,\Phi) + \epsilon_{F}^{T=T_j}(h,\Phi) + B_{\Phi}\text{IPM}_G(p_{\Phi}^{T=T_i}, p_{\Phi}^{T=T_j})),
\end{aligned}
\end{equation*}
where we omit the term of minus variance in the righthand side. And $\epsilon_{F}^{T}(h,\Phi) = \int_{\mathcal{X}}{\ell_{h,\Phi}(x,T)p^{T}(x)dx}$, which represents the learnable factual loss of treatment $T$.
\end{lemma}

In the binary case, Shalit \emph{et al.} separate the regression loss of treatment/non-treatment for the categorical treatments. In our scenario, the treatments are continuous advertising clicks and should be generalizable across different treatments. In this case we define the regression loss with respect to treatments $\{T_i, T_j\}$ as
\begin{equation}\label{eqn:factual_loss}
\epsilon_{i,j}(h,\Phi) = \epsilon_{F}^{T=T_i}(h,\Phi) + \epsilon_{F}^{T=T_j}(h,\Phi).
\end{equation}
Then we are ready to decompose the PEHE error.

\begin{lemma}\label{lemma:pehe_loss_bound}
In the continuous and transitive treatments scenario, it satisfies that
\begin{equation}\label{eqn:pehe_decomposition}
\epsilon_{\text{PEHE}}(f) \le \sum\limits_{i=1}^{n-1}{\int_{\mathcal{X}}{\tau_{i,i+1}^2(x)p(x)dx}}.
\end{equation}
\begin{proof}
It follows that:
\begin{equation*}
\begin{aligned}
\tau_{i,j}(x) &= \hat{\alpha}_{i,j}(x) - \alpha_{i,j}(x) \\
&= f(x,T_j) - f(x,T_i) - m_j(x) + m_i(x) \\
&= f(x,T_j) - f(x, T_{j-1}) + f(x, T_{j-1}) - f(x, T_{j-2}) \\
&+ ... + f(x, T_{i+1}) - f(x, T_i) \\
&- m_j(x) + m_{j-1}(x) - ... - m_{i+1}(x) + m_i(x) \\
&= \hat{\alpha}_{j-1,j}(x) + ... +  \hat{\alpha}_{i,i+1}(x) \\
&- \alpha_{j-1,j}(x) - ... - \alpha_{i,i+1}(x) \\
&= \tau_{i,i+1}(x) + \tau_{i+1,i+2}(x) + ... + \tau_{j-1,j}(x)
\end{aligned}
\end{equation*}
Therefore, according to the Cauchy–Schwarz inequality, we have that
\begin{equation*}
\tau_{i,j}^2(x) \le (j-i)[\tau_{i,i+1}^2(x) + ... + \tau_{j-1,j}^2(x)], \forall j > i.
\end{equation*}
Apparently, $\tau_{i,j}(x) = -\tau_{j,i}(x)$. Then the PEHE loss can then be written as:
\begin{equation*}
\begin{aligned}
\epsilon_{\text{PEHE}}(f) &\le \frac{2}{n(n-1)}\sum\limits_{i=1}^{n}{\sum\limits_{j>i}^{n}{\int_{\mathcal{X}}{\tau_{i,j}^2(x)p(x)dx}}} \\
&= \sum\limits_{i=1}^{n-1}{\int_{\mathcal{X}}{\tau_{i,i+1}^2(x)p(x)dx}}. 
\end{aligned}
\end{equation*}
\end{proof}
\end{lemma}

Combining Lemma \ref{lemma:binary_loss}, Lemma \ref{lemma:pehe_loss_bound} and Eqn. (\ref{eqn:factual_loss}), we obtain a learnable upper bound for PEHE.
\begin{theorem}\label{theorem:pehe_bound}
Under the assumptions in Lemma \ref{lemma:binary_loss} and Lemma \ref{lemma:pehe_loss_bound}, we have that:
\begin{equation}
\epsilon_{\text{PEHE}}(f) \le 2\sum\limits_{i=1}^{n-1}{[\epsilon_{i,i+1}(h,\Phi) + B_{\Phi}\text{IPM}_G(p_{\Phi}^{T=T_i}, p_{\Phi}^{T=T_{i+1}})]}.
\end{equation}
\end{theorem}

Theorem \ref{theorem:pehe_bound} directly points out an algorithm for learning IAE. Relying on continuous and transitive treatments, it generalizes the binary treatments to the multiple treatments advertising scenario. Note that we can compare arbitrary pairwise treatment effect via customized definition of PEHE loss.

\subsection{Algorithm implementation}\label{sec:algorithm}

In the advertising data, the context $x$ mainly embeds the status of an AD with its features in major traffic channels. In our case, we choose the following features to characterize an AD before applying treatment $T$:
\begin{enumerate}
  \item [$\bullet$] IDs: including the commodity ID, the shop ID, the category ID and other one-hot encoding features such as weekdays/weekends etc; 
  \item [$\bullet$] PV sources of the AD in the last day: we aggregate the advertising effects from the major online PV sources of the last day, mainly including sponsored search, recommendations, organic search, etc. Moreover, we also include the time when the AD was created and shelved;
  \item [$\bullet$] PV sources of the AD in the last week: the exponential decaying average advertising effects of last week, including major online PV sources as the above one;  
  \item [$\bullet$] Shop features: including the number of impressions and clicks of the shop in the last day. Moreover, the count of total clicks of the shop, the total number of ADs and also the ADs in sponsored search advertising campaign of the shop are also included; 
  \item [$\bullet$] Competition of the last day: the average ranking of the AD and shop in the bidding process of the last day, with the logarithm of the two values included.
\end{enumerate}
We believe the above features cover most of the data we can fetch online for an AD in Taobao platform.

We learn to infer IAE by minimizing the upper bound of the nominal PEHE loss, using the following objective:
\begin{equation}\label{eqn:objective}
\begin{aligned}
\min\limits_{h,\Phi} \quad & {\frac{2}{N}\sum\limits_{i=1}^{N}{w_i \cdot L(h(\Phi(x_i),t_i), y_i)} + \lambda \cdot \mathscr{R}(h)} \\
&- \frac{\mu_1}{N} \cdot \sum\limits_{i=1}^{N}{L(h(\Phi(x_i),t_i), y_i)\mathbf{1}_{t_i=T_1}} \\
&- \frac{\mu_n}{N} \cdot \sum\limits_{i=1}^{N}{L(h(\Phi(x_i),t_i), y_i)\mathbf{1}_{t_i=T_n}} \\
&+ \beta \cdot \sum\limits_{i=1}^{n-1}{\text{IPM}_G(p_{\Phi}^{T=T_i}, p_{\Phi}^{T=T_{i+1}})}, \\
\text{with} \  & \mu_j = \frac{N_j}{N}, w_i = \mu_{t_i}, N_j = \sum\limits_{i=1}^{N}{\mathbf{1}_{T_j=t_i}}, j=1,...,n, \\
\text{and} \  & \mathscr{R}\  \text{is a model complexity term}.
\end{aligned}
\end{equation}
Note that $w_j,j=1,...,n$ corresponds to the proportion of units applying treatment $T_j$ in the whole population, which is approximated by the sample population. And $\mathbf{1}_{\text{condition}}$ is an indicator function which yields $1$ when the condition is true; otherwise $0$. We use the same $B_{\Phi}=\beta$ across all the IPM distance since they have the same importance in the PEHE definition. We train the models by using stochastic gradient descent to minimize (\ref{eqn:objective}) with $\ell_2$-regularization and $1$-Lipschitz function family $G$. Samples belonging to different treatments share a common representation and hypothesis as shown in Fig. \ref{fig:lvr_model}. The details of the training process is shown in Algorithm \ref{algo:lvr_learning}. The algorithm structure is much like that in \cite{shalit2017estimating} with differences on objective function and gradients. We put it here for completeness.

\begin{algorithm}[h]
\caption{Learning Individual Advertising Effect}\label{algo:lvr_learning}
\begin{algorithmic}
\REQUIRE Samples $\{(x_i,t_i,y_i)\}_{i=1}^{N}$, loss function $L(\cdot)$, regularization factor $\lambda$ and $\beta$, representation network $\Phi_{\mathbf{W}}$ with initial weights $\mathbf{W}$, hypothesis network $h_{\mathbf{V}}$ with initial weights, function family $G$ for IPM distance, regularization norm $\mathscr{R}$.
\ENSURE Representation and hypothesis network.
\STATE Compute $N_j = \sum\limits_{i=1}^{N}{\mathbf{1}_{T_j=t_i}}$\;
\STATE Compute $\mu_j = \frac{N_j}{N}, w_{i} = \mu_{t_i}$\;
 \WHILE{not converged}{
  \STATE Sample mini-batch ${i_1, i_2, ..., i_l} \subset \{1,2,...,N\}$\;
  \STATE Calculate the gradient of the IPM sum term:
  \STATE $g_1 = \nabla_{\mathbf{W}}{\sum\limits_{i=1}^{n-1}{{\text{IPM}_G(p_{\Phi}^{T=T_i}, p_{\Phi}^{T=T_{i+1}})}}}$\;
  \STATE Calculate the gradients of the empirical loss:
  \STATE $g_2 = \nabla_{\mathbf{V}}{L}$, $g_3 = \nabla_{\mathbf{W}}{L}$\;
  \STATE where $L = \frac{2}{m}\sum_j{w_{i_j}\cdot L(h(\Phi(x_{i_j}),t_{i_j}), y_{i_j})} - \frac{\mu_1}{m} \cdot \sum\limits_{i=1}^{N}{L(h(\Phi(x_{i_j}),t_{i_j}), y_{i_j})\mathbf{1}_{t_{i_j}=T_1}} - \frac{\mu_n}{m} \cdot \sum\limits_{i=1}^{N}{L(h(\Phi(x_{i_j}),t_{i_j}), y_{i_j})\mathbf{1}_{t_{i_j}=T_n}}$\;
  \STATE Calculate step size scalar or matrix $\eta$ with Adam \cite{kingma2014adam}\;
  \STATE $[\mathbf{W}, \mathbf{V}] \leftarrow [\mathbf{W} - \eta(\beta g_1 + g_3), \mathbf{V} - \eta(g_2 + 2\lambda \mathbf{V})]$
  \STATE Check convergence condition\;
 }
 \ENDWHILE
\end{algorithmic}
\end{algorithm}

\section{Causal Inference Based Bidding}\label{sec:optimze_plc}

Real-time bidding has been investigated thoroughly in recent years \cite{zhu2017optimized,perlich2012bid,zhang2014optimal}. For e-commercial advertisers in pursuit of conversions, the optimal bidding algorithms share a common value-based form as:
\begin{equation}
bid = \gamma * cvr * ip,
\end{equation}
where $cvr$, $ip$ are the predicted conversion rate and item price, respectively. $\gamma$ is a used to regulate the \emph{return-on-investment (ROI)}/budget of the advertiser. Larger $\gamma$ leads to lower ROI but obtains more impressions. In Taobao sponsored search practice, $\gamma$ is interpreted as the inverse of the expected ROI of the advertiser, which can be estimated from the historical auction log and advertiser's keyword-level bid settings. Apparently, the above bidding only considers the promotion value induced by the advertising PVs, i.e., the direct returns. To bid for the overall returns, we define an AD-level \emph{leverage rate} ($lvr$) based on IAE.
\begin{definition}
The \emph{leverage rate} for context $x$ with advertising clicks changing from $s$ to $t$ is:
\begin{equation}\label{eqn:lvr_definition}
\sigma_{s,t}(x) = \frac{\hat{\alpha}_{s,t}(x)}{t - s}.
\end{equation}
\end{definition}

$lvr$ reflects the average number of all-channel clicks obtained per advertising click invested. Apparently, in the perspective of an AD, $lvr$ is changing as time evolves, which might be influenced by the shifting marketing environment. In the slowly drifting bidding environment, the number of advertising clicks an AD can obtain might also gradually change, which gives an opportunity for the AD to take full advantage of the change. In this sense, $lvr$ can also be seen as the partial derivative of the overall clicks with respect to the advertising clicks. Specifically, for an AD in context $x$, we define the \emph{nominal $lvr$ by taking $s$ and $t$ to be the most recent daily advertising clicks obtained by the same AD}. For example, let $t$ be the number of advertising clicks obtained by the same AD yesterday while $s$ corresponds to that obtained the most recent day other than yesterday, satisfying $s \neq t$. Then the AD-level nominal $lvr$ for today is $\sigma$. We incorporate the nominal $lvr$ in the bidding equation as:
\begin{equation}\label{eqn:lvr_bidding}
bid = \sigma * \gamma * cvr * ip.
\end{equation}

In light of $lvr$, the bidding takes into account the average overall return including both clicks from advertising PV itself and the indirect clicks caused by the advertising effect. In the bidding process, the new bidding formula as in Eqn. (\ref{eqn:lvr_bidding}) should allocate more budget to those with the potential to leverage more all-channel clicks. The overall returns of advertising should be improved given the same budget.

\section{Online Experiments}\label{sec:experiment}

Different from classical machine learning tasks, to evaluate causal effect is intractable due to the missing counterfactual outcomes in reality. Existing binary-treatment work relies on synthetic or simulated dataset such as IHDP and Jobs \cite{yao2018representation} to evaluate. For multiple treatment scenario, to the best of our knowledge, there only exists a multiple intervention breast cancer dataset \cite{yoon2017discovery}, which is not public however. To this end, we turn to evaluate the causal effect estimation by applying it to the online bidding engine in Taobao sponsored search, to compare the bidding performance with the existing online bidding algorithm, which has already been proved to be a strong baseline in practice.

In Taobao sponsored search system, we randomly choose a set of ADs $\mathcal{A}$ to carry out the experiment. The nominal $lvr$ distribution in $\mathcal{A}$ is similar with that in the whole set. We apply the $lvr$-induced bidding for ADs in $\mathcal{A}$ and keep the remaining as the control group. For AD $a_i \in \mathcal{A}$, the bidding equation is:
\begin{equation*}
\begin{aligned}
bid_i &= \kappa * \frac{\sigma_i}{\bar{\sigma}}*\gamma*cvr*ip, \forall a_i \in \mathcal{A},
\bar{\sigma} = \frac{\sum_{i}{\sigma_i}}{|\mathcal{A}|}
\end{aligned}
\end{equation*}
where $\sigma_i$ corresponds to the nominal $lvr$ of AD $a_i$ and $|\mathcal{A}|$ is the cardinality of $\mathcal{A}$. Furthermore, $\kappa$ is utilized to adjust the bidding formula to ensure that the total advertising cost of all the ADs in $\mathcal{A}$ is approximately the same with that applying the existing bidding equation. $\kappa$ is updated daily by an offline replay system and applied online in the new day. The replay system evaluates the cost with respect to bidding given the daily auction log. Note that the same advertising cost implies that the number of advertising clicks is also nearly the same. 

\begin{figure}
\includegraphics[width=3.5in]{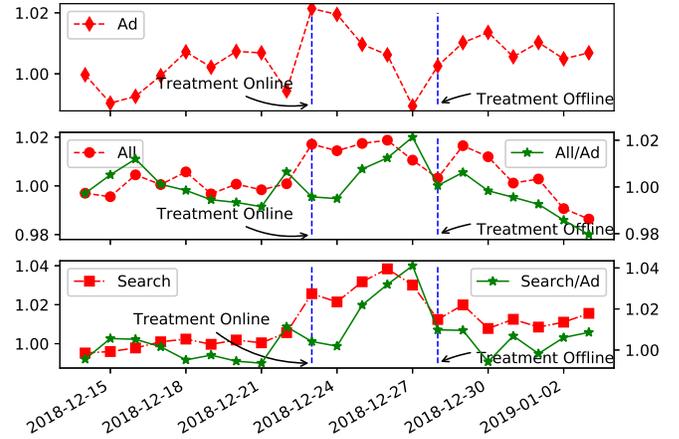}
\caption{$lvr$-bidding performance. The red dashed lines in the upside, middle, and downside plots show the number of advertising clicks (Ad), all-channel clicks excluding advertising clicks (All), and organic search clicks (Search) obtained by ADs in $\mathcal{A}$, respectively. Meanwhile, the green solid line in the middle (All/Ad)/downside (Search/Ad) figure displays the ratio of all-channel clicks (excluding advertising clicks)/organic search clicks over advertising clicks. For commercial secrets, all the values displayed are the relative ratio divided by the mean of the corresponding values before treatment goes online, i.e., from Dec. 14, 2018 to Dec. 22, 2018 (included). The middle and downside figures are equipped with two y-axes, with left for the red line and right for the green line.}\label{fig:increase}
\end{figure}

We compare the number of all-channel clicks obtained in the $lvr$-bidding group, with those of the control group as the baseline. Specifically, we separately show the number of clicks obtained in Taobao organic search engine, which is the major source of free clicks. For commercial secrets, we hide the absolute value but show the relative incremental ratio based on the control group. We display the relative incremental ratio of number of advertising clicks (Ad)/all-channel clicks (All)/organic search clicks (Search) of the $lvr$-bidding group in Fig. \ref{fig:increase}.

The $lvr$-bidding goes online on Dec. 23, 2018 and offline at the end of Dec. 28, 2018 (included). It can be seen that as the $lvr$-bidding goes online, although the number of advertising clicks goes down $2$ percent, the number of all-channel clicks obtained by the same ADs sees an increase of $2$ percent, while the number of organic search clicks increases $4$ percent, which is significant improvement in a giant system like Taobao sponsored search. Furthermore, the ratio of all-channel clicks/organic search clicks over advertising clicks also experiences nearly the same increase pattern. The contradicted variation between advertising and performance here means that advertisers actually improve their marketing performance even with less investments via allocating budgets based on causal effects. As the $lvr$-bidding goes offline, the performance resembles that before the experiment, which shows a relative stable marketing environment during the whole period.

Causal inference-based bidding leads to a clever budget allocation paradigm based on potential return. Such a paradigm is applicable in a lot of e-commerce scenarios such as new product promotion and optimizing the budget allocation in the product life cycle based on the evolving $lvr$.

\section{Conclusions}\label{sec:conclusion}

Online advertising has been the major promotion approach for e-commercial advertisers. However, a reasonable evaluation of both direct and indirect returns induced by advertising has been ignored for a long time. In this paper, we model the relation between advertising investment and return as a causal inference problem with multiple treatments available. Relying on the continuous and transitive treatments, we derive a theoretical upper bound for the expected estimation error of the individual treatment effect. Then a deep representation and hypothesis network is designed to balance the selection bias in causal inference and learn the individual treatment effect. We evaluate the effectiveness of the estimation algorithm by applying it to an industry-level bidding engine, which shows that the causal inference-based bidding outperforms the existing online bidding algorithm. We believe that the proposed bidding paradigm provides new source of performance growth in e-commercial advertising.

In the future, we might infer the causal effect of the more fine-grained PV-level context. Furthermore, the causal effect might be used in the platform level to influence the ranking of ADs for more efficient overall performance.

\bibliographystyle{named}
\bibliography{ijcai19}

\begin{thebibliography}{}

\bibitem[\protect\citeauthoryear{Atan \bgroup \em et al.\egroup
  }{2018}]{atan2018deep}
Onur Atan, James Jordon, and Mihaela van~der Schaar.
\newblock Deep-treat: Learning optimal personalized treatments from
  observational data using neural networks.
\newblock In {\em Thirty-Second AAAI Conference on Artificial Intelligence},
  2018.

\bibitem[\protect\citeauthoryear{Athey and Imbens}{2016}]{athey2016recursive}
Susan Athey and Guido Imbens.
\newblock Recursive partitioning for heterogeneous causal effects.
\newblock {\em Proceedings of the National Academy of Sciences},
  113(27):7353--7360, 2016.

\bibitem[\protect\citeauthoryear{Austin}{2011}]{austin2011introduction}
Peter~C Austin.
\newblock An introduction to propensity score methods for reducing the effects
  of confounding in observational studies.
\newblock {\em Multivariate behavioral research}, 46(3):399--424, 2011.

\bibitem[\protect\citeauthoryear{Bottou \bgroup \em et al.\egroup
  }{2013}]{bottou2013counter}
L\'{e}on Bottou, Jonas Peters, Joaquin~Qui\ {n}onero Candela, Denis~X. Charles,
  D.~Max Chickering, Elon Portugaly, Dipankar Ray, Patrice Simard, and
  Ed~Snelson.
\newblock Counterfactual reasoning and learning systems: The example of
  computational advertising.
\newblock {\em Journal of Machine Learning Research}, 14:3207--3260, 2013.

\bibitem[\protect\citeauthoryear{Dalessandro \bgroup \em et al.\egroup
  }{2012}]{dalessandro2012causally}
Brian Dalessandro, Claudia Perlich, Ori Stitelman, and Foster Provost.
\newblock Causally motivated attribution for online advertising.
\newblock In {\em Proceedings of the Sixth International Workshop on Data
  Mining for Online Advertising and Internet Economy}, page~7. ACM, 2012.

\bibitem[\protect\citeauthoryear{Diemert \bgroup \em et al.\egroup
  }{2017}]{diemert2017attribution}
Eustache Diemert, Julien Meynet, Pierre Galland, and Damien Lefortier.
\newblock Attribution modeling increases efficiency of bidding in display
  advertising.
\newblock In {\em Proceedings of the ADKDD'17}, page~2. ACM, 2017.

\bibitem[\protect\citeauthoryear{Edquid}{2016}]{edquid17}
R.~Edquid.
\newblock 10 of the largest ecommerce markets in the world by country.
\newblock
  \url{https://www.business.com/articles/10-of-the-largest-ecommerce-markets-in-the-world-b/},
  2016.
\newblock Accessed Jan. 22, 2018.

\bibitem[\protect\citeauthoryear{Johansson \bgroup \em et al.\egroup
  }{2016}]{johansson2016learning}
Fredrik Johansson, Uri Shalit, and David Sontag.
\newblock Learning representations for counterfactual inference.
\newblock In {\em International Conference on Machine Learning}, pages
  3020--3029, 2016.

\bibitem[\protect\citeauthoryear{Kingma and Ba}{2014}]{kingma2014adam}
Diederik~P Kingma and Jimmy Ba.
\newblock Adam: A method for stochastic optimization.
\newblock {\em arXiv preprint arXiv:1412.6980}, 2014.

\bibitem[\protect\citeauthoryear{Lopez \bgroup \em et al.\egroup
  }{2017}]{lopez2017estimation}
Michael~J Lopez, Roee Gutman, et~al.
\newblock Estimation of causal effects with multiple treatments: a review and
  new ideas.
\newblock {\em Statistical Science}, 32(3):432--454, 2017.

\bibitem[\protect\citeauthoryear{Perlich \bgroup \em et al.\egroup
  }{2012}]{perlich2012bid}
Claudia Perlich, Brian Dalessandro, Rod Hook, Ori Stitelman, Troy Raeder, and
  Foster Provost.
\newblock Bid optimizing and inventory scoring in targeted online advertising.
\newblock In {\em Proceedings of the 18th ACM SIGKDD international conference
  on Knowledge discovery and data mining}, pages 804--812. ACM, 2012.

\bibitem[\protect\citeauthoryear{Peysakhovich and
  Lada}{2016}]{peysakhovich2016combining}
Alexander Peysakhovich and Akos Lada.
\newblock Combining observational and experimental data to find heterogeneous
  treatment effects.
\newblock {\em arXiv preprint arXiv:1611.02385}, 2016.

\bibitem[\protect\citeauthoryear{Rubin}{2005}]{rubin2005causal}
Donald~B Rubin.
\newblock Causal inference using potential outcomes: Design, modeling,
  decisions.
\newblock {\em Journal of the American Statistical Association},
  100(469):322--331, 2005.

\bibitem[\protect\citeauthoryear{Shalit \bgroup \em et al.\egroup
  }{2017}]{shalit2017estimating}
Uri Shalit, Fredrik~D Johansson, and David Sontag.
\newblock Estimating individual treatment effect: generalization bounds and
  algorithms.
\newblock In {\em International Conference on Machine Learning}, pages
  3076--3085, 2017.

\bibitem[\protect\citeauthoryear{Taddy \bgroup \em et al.\egroup
  }{2016}]{taddy2016nonparametric}
Matt Taddy, Matt Gardner, Liyun Chen, and David Draper.
\newblock A nonparametric bayesian analysis of heterogenous treatment effects
  in digital experimentation.
\newblock {\em Journal of Business \& Economic Statistics}, 34(4):661--672,
  2016.

\bibitem[\protect\citeauthoryear{Wager and Athey}{2017}]{wager2017estimation}
Stefan Wager and Susan Athey.
\newblock Estimation and inference of heterogeneous treatment effects using
  random forests.
\newblock {\em Journal of the American Statistical Association},
  (just-accepted), 2017.

\bibitem[\protect\citeauthoryear{Wilkens \bgroup \em et al.\egroup
  }{2017}]{wilkens2017gsp}
Christopher~A Wilkens, Ruggiero Cavallo, and Rad Niazadeh.
\newblock Gsp: the cinderella of mechanism design.
\newblock In {\em Proceedings of the 26th International Conference on World
  Wide Web}, pages 25--32. International World Wide Web Conferences Steering
  Committee, 2017.

\bibitem[\protect\citeauthoryear{Xu \bgroup \em et al.\egroup
  }{2016}]{xu2016lift}
Jian Xu, Xuhui Shao, Jianjie Ma, Kuang-chih Lee, Hang Qi, and Quan Lu.
\newblock Lift-based bidding in ad selection.
\newblock In {\em AAAI}, pages 651--657, 2016.

\bibitem[\protect\citeauthoryear{Yao \bgroup \em et al.\egroup
  }{2018}]{yao2018representation}
Liuyi Yao, Sheng Li, Yaliang Li, Mengdi Huai, Jing Gao, and Aidong Zhang.
\newblock Representation learning for treatment effect estimation from
  observational data.
\newblock In {\em Advances in Neural Information Processing Systems}, pages
  2638--2648, 2018.

\bibitem[\protect\citeauthoryear{Yoon \bgroup \em et al.\egroup
  }{2017}]{yoon2017discovery}
Jinsung Yoon, Camelia Davtyan, and Mihaela van~der Schaar.
\newblock Discovery and clinical decision support for personalized healthcare.
\newblock {\em IEEE journal of biomedical and health informatics},
  21(4):1133--1145, 2017.

\bibitem[\protect\citeauthoryear{Zhang \bgroup \em et al.\egroup
  }{2014}]{zhang2014optimal}
Weinan Zhang, Shuai Yuan, and Jun Wang.
\newblock Optimal real-time bidding for display advertising.
\newblock In {\em Proceedings of the 20th ACM SIGKDD international conference
  on Knowledge discovery and data mining}, pages 1077--1086. ACM, 2014.

\bibitem[\protect\citeauthoryear{Zhang}{2016}]{zhang2016optimal}
Weinan Zhang.
\newblock {\em Optimal real-time bidding for display advertising}.
\newblock PhD thesis, UCL (University College London), 2016.

\bibitem[\protect\citeauthoryear{Zhu \bgroup \em et al.\egroup
  }{2017}]{zhu2017optimized}
Han Zhu, Junqi Jin, Chang Tan, Fei Pan, Yifan Zeng, Han Li, and Kun Gai.
\newblock Optimized cost per click in taobao display advertising.
\newblock In {\em Proceedings of the 23rd ACM SIGKDD International Conference
  on Knowledge Discovery and Data Mining}, pages 2191--2200. ACM, 2017.

\end{thebibliography}

\end{document}